\documentclass[english,twocolumn,prl,showpacs]{revtex4}
\usepackage[latin1]{inputenc}
\usepackage{amsmath}
\usepackage{graphicx}

\batchmode

\newcommand{\beq}{\begin{equation}}
\newcommand{\eeq}{\end{equation}}

\usepackage{babel}
\makeatother

\begin{document}

\title{Thermal equilibrium of two quantum Brownian particles}

\author{D. M. Valente and A. O. Caldeira}

\affiliation{Departamento de F\'{\i}sica da Mat{\'e}ria Condensada, Instituto de F\'{\i}sica
Gleb Wataghin, Universidade Estadual de Campinas, CEP 13083-970, Campinas-SP,
Brazil}

\begin{abstract}
The influence of the environment in the thermal equilibrium
properties of a bipartite continuous variable quantum system is
studied. The problem is treated within a system-plus-reservoir
approach. The considered model reproduces the conventional Brownian motion when
the two particles are far apart and induces an effective
interaction between them, depending on the choice of the spectral
function of the bath. The coupling between the system and the
environment guarantees the translational invariance of the system
in the absence of an external potential. The entanglement between
the particles is measured by the logarithmic negativity, which is
shown to monotonically decrease with the increase of the
temperature. A range of finite temperatures is found in which
entanglement is still induced by the reservoir.
\end{abstract}


\maketitle

\section{I. Introduction}

The classical Brownian motion has become one of the most important
paradigms of irreversible processes in physics, where
out-of-equilibrium fluctuations play an important role.  The
equation that describes such an open system is the famous Langevin
equation, in which dissipation and noise are present.

However, when an open system is treated  quantum mechanically, the
problem has to be analyzed much more carefully. Since quantum
mechanics is designed for dealing with hamiltonian  systems, one
is bounded to employ the so-called system-plus-reservoir approach
to tackle this problem. Basically, one can consider the system and
the environment together as an isolated system and quantize it. It
is also possible, depending on the conditions to be obeyed
\cite{annals,Hedegard,solitons}, modelling the interaction between
the reservoir and the system in an appropriate way and consider
the environment as a bath of non-interacting harmonic oscillators.
Once this is done it is  possible to reproduce the  Langevin
equation for the variable which describes the dynamics of the
system of interest in the adequate limits \cite{physica}. In this
context of open quantum systems one can describe  dissipation
\cite{annals} and the loss of quantum mechanical coherence which
is known in the literature as decoherence \cite{decor}.

When the interest is in the dynamics of two Brownian particles in
a common reservoir, some changes have to be made in the
traditional bath-of-oscillators model. Those changes are described
in Ref.\cite{Duarte}, where the classical equations of motion for
two Brownian particles have been obtained. The same generalization
of the model has been used to evaluate the quantum dynamics of two
Brownian particles \cite{Oscar}. It is now our task to study the
thermal equilibrium properties of this system.

At first sight, one could expect that only dissipation and
decoherence should appear in this problem. But, under certain
conditions, the reservoir can correlate these two particles, an
effect similar to  the Cooper pairing in BCS superconductors
\cite{BCS}. In particular, in the model of Ref.\cite{Duarte}, an
effective potential between the particles appears as a consequence
of the hypothesis of the model. A possible  correlation effect
induced by this potential is the entanglement, a property of
quantum states of great interest to the field of quantum
information \cite{Adesso, Simon, Peres}. We are going to quantify
the entanglement between the particles induced by the reservoir as
a function of the temperature of the composite system. Some
attempts to address this question have already been made
\cite{Horhammer, J Paz}, but we believe that  the present model
\cite{Duarte, Oscar} is more adequate to account for the correct
quantum behavior of two particles in the common heat bath.

In section II we introduce the generalized model in some detail.
The classical equations of motion are also reviewed. Then, we
evaluate the equilibrium reduced density matrix for two quantum
Brownian particles. In section III, we use this density matrix to
understand how the entanglement between the particles is related
to the temperature of the environment.

\section{II. Model and equilibrium density operator}

As we mentioned before, the possibility of modelling a Brownian
particle within the system-plus-reservoir approach is very
well-known. In particular, if the environment is  represented by a
bath of non-interacting harmonic oscillators  whose coordinates
are coupled bilinarly to the coordinate of the particle of
interest one can reproduce the Langevin equation in the classical
limit \cite{annals,physica}.

However, depending on the specific coupling between the system of
interest and the reservoir, a  generalization of that conventional
model is needed for the case of two brownian particles
\cite{Duarte}. In what follows we review the important aspects of
such a model for two particles in a common reservoir and present
our main results for their equilibrium density matrix. This can be
compared to the dynamical case \cite{Oscar} which has been recently
studied.

We follow the Feynman path integral formalism, which implies that
the action of the reservoir on the system of interest can be
described in terms of the so-called  influence functional
\cite{Feynman-Vernon}. In the equilibrium case it is actually an
Euclidean version thereof to which for simplicity we still refer
as the influence functional.

As an intermediate step, we show that the influence functional of
two particles can be separated in three terms: one for each
independent subsystem and another one for the effective interaction
between them. It is enlightening to notice that indeed the
influence functional for each subsystem within the present method
reproduces the result obtained by the bilinear (conventional)
coupling model.

\subsection{A. Model and classical equations of motion}

The Lagrangian of the composite system reads

\beq
L = L_S + L_I + L_R.
\label{lag}\eeq

The term $L_S$ is the Lagrangian for the system of interest, $L_I$
is the one for the interaction between the system and the
reservoir and $L_R$ is the Lagrangian for the bath. Explicitly, we
have for two free particles

\beq
L_S = \frac{1}{2}M\dot{x}_1^2+\frac{1}{2}M\dot{x}_2^2.
\eeq

The reservoir will be described by a symmetrized collection of independent harmonic modes,

\beq
L_R = \frac{1}{2}\sum_{k=1}^N m_k \left(\dot{R}_k\dot{R}_{-k}-\omega_k^2R_kR_{-k}\right).
\eeq

A non-linear coupling in the coordinates of the system appears in
the interaction Lagragian  whose form is

{\footnotesize\beq
L_I = -\frac{1}{2} \sum_k \left\{ [C_{-k}(x_1)+C_{-k}(x_2)]R_k
+ [C_{k}(x_1)+C_{k}(x_2)]R_{-k}\right\}.
\eeq}

The local interaction of a particle with a spatially homogeneous
environment can be represented by a function of the type

\beq
C_{k}(x) = \kappa_k e^{ikx}.
\label{cedeka}\eeq

Eq. (\ref{cedeka}) appears, for example, in the polaron problem
\cite{Weiss}. This function is capable of preserving the
translational invariance of the system in the absence of an
external potential.

The properties of this model as well as its suitability to
describe an effective Brownian motion are deeply discussed in
\cite{Duarte} and \cite{Oscar}. In particular, reference
\cite{Duarte} shows the classical equations of motion for

\beq \zeta \equiv \frac{x_1 + x_2}{2}  \ \ \ \mbox{and} \ \ \ \xi
\equiv x_1 - x_2, \label{coordtransf}\eeq which are respectively
the center of mass and relative coordinates of the two particles.
Those equations are

\begin{equation}
M\ddot{\xi}(t) + (\eta - \tilde{\eta}[\xi(t)])\dot{\xi}(t)+ V_{eff}'(\xi(t)) = F_\xi(t)
\end{equation}
and
\begin{equation}
M\ddot{\zeta}(t) + (\eta + \tilde{\eta}[\zeta(t)])\dot{\zeta} = F_\zeta(t),
\end{equation}
with
\begin{equation}
\tilde{\eta}[\xi(t)] \equiv \eta\left(
\frac{1}{(k_0^2\xi^2(t)+1)^2} -
\frac{4k_0^2\xi^2(t)}{(k_0^2\xi^2(t)+1)^3}  \right)
\label{eta}\end{equation} and \beq V_{eff}(\xi(t))\equiv
-\frac{2\Omega\eta}{\pi
k_0^2}\left(\frac{1}{k_0^2\xi^2(t)+1}\right). \label{veff}\eeq We
define $\eta = \sum_k k^2 \kappa_k\kappa_{-k}f(k)$ as the
dissipation constant. The other constants are $\Omega$, a cutoff
frequency for the oscillators of the bath, and $k_0$, the inverse
of its characteristic length scale. In fermionic environments, for
example, this length is related to the Fermi wave number $k_F$
\cite{Hedegard,Guinea}. The functions $F_\xi(t)$ and $F_\zeta(t)$
are fluctuating forces depending on the initial conditions
\cite{Duarte} imposed to the composite system.

Note that (\ref{eta}) and (\ref{veff}) introduce some new
phenomena into this problem. The former indicates how the net
dissipation depends on the distance between the particles whereas
the  latter presents an effective potential between the particles
mediated by the environment. In a regime in which the particles
are close enough ($\xi << k_0^{-1}$), the center of mass still
behaves like a Brownian particle but the relative coordinate
describes an undamped harmonic oscillator of frequency $\omega_0$,
with $\omega_0^2 \equiv 4\eta\Omega/M\pi$. When the two particles
are far apart ($\xi >> k_0^{-1}$) both variables describe standard
Brownian motion with constant damping $\eta$.

\subsection{B. Two particle equilibrium density operator}

The total lagrangian in Eq. (\ref{lag}) generates a hamiltonian $H$ for the
composite system that can be used to describe the density operador of
the system of interest $\rho_S$,

\begin{equation}
\rho_S = \mbox{Tr}_R[\rho_{tot}],
\end{equation}
where
\begin{equation}
\rho_{tot} = \mathcal{Z}^{-1} e^{-\beta H},
\end{equation}
such that $\mathcal{Z}$ is a normalization constant (the partition
function of the full system) and $\beta$ satisfies $\beta = (k_B
T)^{-1}$ where $k_B$ is the Boltzman constant and $T$, the
temperature, as usual.

In the coordinate representation \cite{Feynman-Vernon,Hedegard},
we get

\beq
\rho_S(x_1,y_1;x_2,y_2,\beta) = \int_{y_1}^{x_1}\mathcal{D}[q_1(\tau)]\int_{y_2}^{x_2}\mathcal{D}[q_2(\tau)]
\nonumber\eeq
\beq
\exp\left(-\frac{1}{\hbar}\left\{S_0^E[q_1(\tau)]+S_0^E[q_2(\tau)]\right\}\right)\times F[q_1(\tau),q_2(\tau)].
\label{rrrooo}\eeq
The term $S_0^E$ is the Euclidean action of the isolated system,

\beq
S_0^E[x(\tau)] = \int_0^{\hbar\beta}\left( \frac{1}{2}M\left( \frac{dx}{d\tau} \right)^2 + V(x) \right) d\tau,
\eeq
with $V(x)$ being an external potential, which is null in our case.
The influence functional is

\begin{equation}
F[q_1(\tau),q_2(\tau)] = \mbox{Tr}_R\left[  \exp\left(-\frac{1}{\hbar}\int_0^{\hbar\beta}H_{IR}(\tau)d\tau\right)  \right],
\label{f}\end{equation}
in which $H_{IR} = H_R + H_I[q_1(\tau)] + H_I[q_2(\tau)]$, with

\begin{equation}
H_I[q_1(\tau)] = \frac{1}{2} \sum_k \left\{C_{-k}[q_1(\tau)]R_k + C_k[q_1(\tau)]R_{-k}\right\},
\label{hium}\end{equation}
\begin{equation}
H_I[q_2(\tau)] = \frac{1}{2} \sum_k \left\{C_{-k}[q_2(\tau)]R_k + C_k[q_2(\tau)]R_{-k}\right\}
\label{hidois}\end{equation}
and
\begin{equation}
H_R = \frac{1}{2} \sum_k \left(\frac{P_kP_{-k}}{2m_k}+\frac{m_k\omega_k^2}{2}R_kR_{-k}\right).
\end{equation}

It is a simple task to show that the influence functional can be
equally written in the interaction picture as

\begin{equation}
F[q_1(\tau),q_2(\tau)] = \mbox{Tr}_R\left[  e^{-\beta H_R} \wp    \right].
\label{inffunc}\end{equation}
The operator $\wp$ defined above satisfies the relation

\beq
\frac{d\wp}{d\beta} = -\left\{\tilde{H}_I[q_1(\beta)] + \tilde{H}_I[q_2(\beta)]\right\}\wp,
\eeq
for which the formal solution is
\beq
\wp \equiv  T \exp\left( -\frac{1}{\hbar}\int_0^{\hbar\beta}\left\{
\tilde{H}_I[q_1(\tau)]+\tilde{H}_I[q_2(\tau)]\right\}d\tau  \right),
\label{adicional}\eeq
where $\tilde{H}_I[q_i(\tau)] \equiv e^{\beta H_R} H_I[q_i(\tau)] e^{-\beta H_R}$ and
$T$ is the temporal ordering operator. In what follows we
explain the reason for using the interaction picture.

A reservoir must be sufficiently robust so that any interaction
with the system of interest is only capable of weakly perturbing
it. Assuming such a condition, we apply an imaginary time
dependent perturbation theory for the bath \cite{Fetter}.
Therefore we can expand (\ref{adicional}) to second order in
perturbation strength, which results in

\begin{equation}
\wp \approx
 {\small
 1 - \frac{1}{\hbar}\int_0^{\hbar\beta} Y(\tau)d\tau
    +\frac{1}{\hbar^2} \int_0^{\hbar\beta}d\tau\int_0^{\tau}d\tau' Z(\tau,\tau')} ,
\label{funfa}\end{equation}

where
\begin{equation}
Y(\tau) \equiv \left\{\tilde{H}_I[q_1(\tau)]+\tilde{H}_I[q_2(\tau)]\right\}
\label{y}\end{equation}
and
\beq
Z(\tau,\tau') \equiv Y(\tau)Y(\tau').
\label{z}\eeq

We can now identify the term $e^{-\beta H_R}$ in Eq.(\ref{inffunc}) with
the equilibrium density operator of the reservoir, $\rho_R = e^{-\beta H_R}$, and
use the notation $\langle\mathcal{O} \rangle = \mbox{Tr}_R[\rho_R \mathcal{O}]$ for
a general operator $\mathcal{O}$.

In this new notation, we see from (\ref{inffunc}) that we have to
evaluate $\langle \wp \rangle$. By (\ref{funfa}), (\ref{y}) and
(\ref{z}) we note that this procedure involves the evaluation of
terms such $\langle Y(\tau) \rangle$ and $\langle Z(\tau, \tau')
\rangle$. Observing the definition (\ref{hium}) we find

\beq
\langle \tilde{H}_I[q_i(\tau)]\rangle = \frac{1}{2}\sum_k \left\{   C_{-k}[q_i(\tau)]   \langle R_k\rangle  +  C_{k}[q_i(\tau)]
   \langle R_{-k}\rangle\right\},
\label{higr}
\eeq
which is identically zero, since $\langle R_k\rangle = 0$ $\forall k$ when the bath is in thermal equilibrium.

This fact implies that the functional in (\ref{inffunc}) can be
equivalently obtained from the expansion of the following
exponential

\beq
F[q_1(\tau),q_2(\tau)] =
\exp\left[  \frac{1}{\hbar^2}\int_0^{\hbar\beta}d\tau\int_0^\tau d\tau'
\left(\sum_{i,j = 1}^2 A_{ij}  \right)  \right],
\label{somadea}
\eeq
where
\begin{equation}
A_{ij} \equiv \langle   H_I[q_i(\tau)]H_I[q_j(\tau')] \rangle.
\label{a}\end{equation} As in (\ref{higr}), the ``tilde'' has
already been dropped because of the cyclic  property of the trace.

Assuming again (\ref{hium}) and (\ref{hidois}), we have
\begin{equation}
A_{ij} = \frac{1}{4}\sum_{k,k'}C_{-k}[q_i(\tau)]C_{-k'}[q_j(\tau')]\langle R_{k}(\tau)R_{k'}(\tau') \rangle
\nonumber\eeq
\beq
+ C_{-k}[q_i(\tau)]C_{k'}[q_j(\tau')]
\langle R_{k}(\tau)R_{-k'}(\tau') \rangle
\nonumber
\end{equation}
\begin{equation}
+C_k[q_i(\tau)]C_{-k'}[q_j(\tau')]\langle
R_{-k}(\tau)R_{k'}(\tau') \rangle \nonumber\eeq \beq +
C_k[q_i(\tau)]C_{k'}[q_j(\tau')] \langle
R_{-k}(\tau)R_{-k'}(\tau') \rangle.
\label{somadupla}\end{equation} With the help of the condition of
invariance of the Lagrangian under translation of the system of
interest, only the terms with $k = -k'$ survive in the sum. Then,
we have

\beq
A_{ij} = \frac{1}{2}\sum_k(C_{-k}[q_i(\tau)]C_{k}[q_j(\tau')]
\nonumber\eeq
\beq
+C_{k}[q_i(\tau)]C_{-k}[q_j(\tau')])\langle R_k(\tau)R_{-k}(\tau') \rangle.
\eeq

The treatment of the correlation function $\Phi_k(\tau-\tau') \equiv \langle R_k(\tau)R_{-k}(\tau') \rangle$ requires
the use of the fluctuation-dissipation theorem \cite{Weiss}, which states that

\begin{equation}
\langle R_k(t)R_{-k}(s) \rangle =
\frac{\hbar}{\pi}\int_{-\infty}^\infty d\omega \
\chi''_k(\omega)\frac{e^{-i\omega(t-s)}}{1-e^{-\omega\hbar\beta}}.
\label{flut-diss}\end{equation} In the above relation we use
$\chi''_k(\omega)$ for the imaginary part of the Fourier transform
of the linear response function of the $k^{th}$ oscillator. This
function can be modelled in terms of the macroscopic behavior of
the bath, since describing its microscopic details is not our
intention here. Such an approach allows us to consider the
harmonic oscillators of the environment as very weakly damped ones
\cite{Duarte}, yielding

\beq \chi''_k(\omega) =
\frac{\gamma_k\omega}{m_k[(\omega^2-\omega_k^2)^2+\omega^2\gamma_k^2]},
\label{chiosc} \eeq which  permits a simple separation of the
length and time scales of the system. The damping constant of each
bath oscillator is given by $\gamma_k$. In the low frequency limit
we have

\beq \chi''_k(\omega) = f(k)\omega \Theta(\Omega-\omega),
\label{xi}\eeq where we have  introduced the same high frequency
cutoff $\Omega$ of the part A of this section. The Markovian
dynamics is achieved when we take the limit $\Omega \rightarrow
\infty$ and the function $f(k)$ represents the nonlocal influence
of the bath. A functional dependence like Eq.(\ref{xi}) for the
dynamical susceptibility of the bath has also been employed in
\cite{Hedegard,Guinea,Duarte}.

In order to calculate $\Phi_k(\tau-\tau')$, we make the usual
analytical extension  $t=-i\tau$ and $s = -i\tau'$ in
(\ref{flut-diss}). As we can see from (\ref{chiosc}) the response
function is odd in frequency, $\chi''_k(\omega) =
-\chi''_k(-\omega)$, which allows us to rewrite (\ref{flut-diss})
as

\beq
\Phi_k(\tau-\tau') = \frac{\hbar}{\pi}\int_0^\infty d\omega \ \chi''_k(\omega) \frac{\cosh\left[ \omega\left(|\tau-\tau'|-\frac{\hbar\beta}{2}\right)\right]}{\sinh \left(\frac{\hbar\omega\beta}{2}\right)}.
\label{phizao}
\eeq

With the correlation function evaluated in terms of the
phenomenological parameters of the problem, we can return to the
influence functional and evaluate it with the choice we made in
(\ref{cedeka}), which leads us to

\begin{equation}
F[q_1(\tau),q_2(\tau)] =
\exp \left[  \frac{1}{\hbar^2}\int_0^{\hbar\beta}d\tau\int_0^\tau d\tau' \
\alpha(\tau,\tau')   \right],
\label{func}\end{equation}
where we define
\beq
\alpha(\tau,\tau') \equiv \sum_k \kappa_k\kappa_{-k} [  \Xi_k(\tau,\tau')   \Phi_k(\tau-\tau')   ]
\eeq
and
\begin{equation}
\Xi_k[\tau, \tau'] \equiv \sum_{i,j = 1}^{2} \cos[k(q_i(\tau) - q_j(\tau'))].
\end{equation}

In the beginning of section II we mentioned that we were going to
show that the influence functional could be separated in three
terms. Moreover, two of them could be interpreted as the
functionals of the subsystems only. Now we make this statement
more clear. Observe that (\ref{func}) can be written in the
product form $F[q_1(\tau),q_2(\tau)] =
F[q_1(\tau)]F[q_2(\tau)]F_{int}[q_1(\tau),q_2(\tau)]$, where we
define

\beq
F[q_{i}(\tau)] \equiv \exp \left[  \frac{1}{\hbar^2} \int_0^{\hbar\beta}d\tau\int_0^\tau d\tau' \
\alpha_i(\tau,\tau')  \right]
\label{efei}\eeq
and
\beq
\alpha_i(\tau,\tau') \equiv \sum_k \kappa_k\kappa_{-k}  \Phi_k(\tau-\tau') \cos[k(q_{i}(\tau)-q_{i}(\tau'))],
\label{alfai}\eeq
for $i = 1, 2$.

We can assume that the trajectory fluctuations of each particle
are restricted to a small region compared to the characteristic
length of the bath, $k_0^{-1}$. Then, we expand the cossine
function to second order in Eq.(\ref{alfai}). The zeroth  order
term will be left for the normalization of the density matrix. To
proceed, we define

\beq K(\tau-\tau') \equiv  \frac{\eta}{\pi} \int_0^\Omega d\omega
\ \omega \frac{\cosh\left[
\omega\left(\tau-\tau'-\frac{\hbar\beta}{2}\right)\right]}{\sinh
\left(\frac{\hbar\omega\beta}{2}\right)} \eeq and identify, with
the help of (\ref{xi}),

\beq
\Phi_k(\tau-\tau') = f(k)\frac{\hbar}{\eta}K(\tau-\tau').
\eeq
Using the fact that $\eta = \sum_k \kappa_k\kappa_{-k}k^2f(k)$, we have

\begin{flushleft}
$F[q_i(\tau)] = $
\end{flushleft}
\beq
=\exp\left[      -\frac{1}{2\hbar}  \int_0^{\hbar\beta}d\tau\int_0^\tau d\tau'
 K(\tau-\tau')(q_i(\tau) - q_i(\tau'))^2      \right].
\label{funcinfump}\eeq
Eq.(\ref{funcinfump}) is exactly the influence functional of one particle
in thermal equilibrium with a bath of harmonic oscillators
with bilinear coupling \cite{Weiss}.

The third term of that product is $F_{int}[q_1(\tau),q_2(\tau)]$,
which involves the correlations between the coordinates of the two
particles. However, such a correlation relates the coordinates in
different imaginary times, in the form
$\cos[k(q_1(\tau)-q_2(\tau'))]$. To deal with this problem we
choose to apply the inverse coordinate transformation of that made
in (\ref{coordtransf}). That is, we write $q_1$ and $q_2$ in terms
of the center of mass $\zeta$ and the relative coordinate $\xi$
and substitute them in $F_{int}[q_1(\tau),q_2(\tau)]$. With some
algebra we get

{\small\beq
F_{int}[\zeta(\tau),\xi(\tau)] =
\exp \left[  \frac{1}{\eta\hbar}\int_0^{\hbar\beta}d\tau\int_0^\tau d\tau' \
\Lambda_{int}(\tau,\tau') K(\tau-\tau')  \right],
\eeq}
where we define
\beq
\Lambda_{int}(\tau,\tau') \equiv \left( \sum_k\kappa_k\kappa_{-k}f(k) \lambda^{(k)}_{int}(\tau,\tau')   \right),
\eeq
with
\begin{equation}
\lambda^{(k)}_{int}(\tau,\tau') \equiv 2 \cos\left[k\left(  \zeta(\tau)-\zeta(\tau') \right)   \right]\cos[k \ \theta(\tau,\tau')]
\label{cosseno}\end{equation}
and
\begin{equation}
\theta(\tau,\tau') \equiv \left(\frac{\xi(\tau)+\xi(\tau')}{2}\right).
\label{plus}\end{equation}

We again evoke the fact that the fluctuations of the center of mass trajectory
are small compared to the length scale of the bath, that is, $\zeta(\tau)\approx \zeta(\tau')$.
So, we expand Eq.(\ref{cosseno}) to second order and get

\begin{equation}
\Lambda_{int}(\tau,\tau') = \ P(\tau,\tau') \ + \ Q(\tau,\tau'),
\end{equation}
where we define $P(\tau,\tau')$ and $Q(\tau,\tau')$ as
\begin{equation}
P(\tau,\tau') \equiv 2 \sum_k\kappa_k\kappa_{-k}f(k)   \cos[k \ \theta(\tau,\tau')]
\end{equation}
and
\begin{equation}
Q(\tau,\tau') \equiv -\sum_k\kappa_k\kappa_{-k}f(k)k^2  \cos[k \ \theta(\tau,\tau')].
\end{equation}

The evaluation of the sums can be done in the same way as in
\cite{Duarte}. Briefly explaining, one can turn the sum into an
integral by introducing the function $g(k) \equiv
\frac{L}{2\pi}\kappa_k\kappa_{-k}f(k)$ and, following
\cite{Duarte}, we can model this function by $g(k) = e^{-k/k_0}$,
where $k_0$ has been already defined. Surprisingly the integrals
can be exactly solved and the result is

\begin{equation}
P(\tau,\tau') = -\frac{\pi}{\Omega} \  V_{eff}(\theta(\tau,\tau'))
\label{u}\end{equation}
and
\begin{equation}
Q(\tau,\tau') = - \tilde{\eta}[\theta(\tau,\tau')],
\label{v}\end{equation} with $\tilde{\eta}[\theta(\tau,\tau')]$
and $V_{eff}(\theta(\tau,\tau'))$ defined in (\ref{eta}) and
(\ref{veff}), respectively. At this point the physical meaning of
the parameters of the model emerges.

If we rewrite also $F[q_1(\tau)]$ and $F[q_2(\tau)]$ in terms of the new
variables, we see that the equilibrium density matrix of the two brownian
particles is

{\small\begin{equation} \rho_S(\zeta_f,\zeta_i;\xi_f,\xi_i,\beta)
= \int_{\zeta_i}^{\zeta_f}\mathcal{D}[\zeta(\tau)]
\int_{\xi_i}^{\xi_f}\mathcal{D}[\xi(\tau)] \exp\left[
-\frac{1}{\hbar} S_{eff} \right], \label{ros}\end{equation}} where
\begin{equation}
S_{eff} \equiv \int_0^{\hbar\beta} d\tau \left[ M\dot{\zeta}^2 +\frac{M}{4}\dot{\xi}^2  \right] + \Gamma
\label{seff}\end{equation}
and
\begin{equation}
\Gamma \equiv \frac{1}{\eta} \int_0^{\hbar\beta}d\tau\int_0^\tau d\tau'\tilde{\eta}[\theta(\tau,\tau')] K(\tau-\tau')(\zeta(\tau)-\zeta(\tau'))^2+
\nonumber\end{equation}
\begin{equation}
+  \int_0^{\hbar\beta}d\tau\int_0^\tau d\tau' K(\tau-\tau')(\zeta(\tau)-\zeta(\tau'))^2+
\nonumber\end{equation}
\begin{equation}
+ \frac{\pi}{\eta\Omega} \int_0^{\hbar\beta}d\tau\int_0^\tau d\tau' K(\tau-\tau')V_{eff}[\theta(\tau,\tau')]+
\nonumber\end{equation}
\begin{equation}
+ \frac{1}{4} \int_0^{\hbar\beta}d\tau\int_0^\tau d\tau' K(\tau-\tau')(\xi(\tau)-\xi(\tau'))^2.
\label{gama}\end{equation}

It is possible to interpret   each term in the equations above in
order to understand the solution's consistency and solve the
Feynman  path integrals.

The center of mass $\zeta$ behaves like a free quantum brownian particle in
thermal equilibrium with a reservoir with dissipation constant $\eta$ and
an ``anomalous'' dissipation given by $\tilde{\eta}[\theta(\tau,\tau')]$.

The relative coordinate $\xi$ behaves like a quantum Brownian
particle in a potential $V_{eff}(\xi)$ in thermal equilibrium with
a reservoir of dissipation constant $\eta$.

Although we have a result that agrees with the qualitative
interpretation induced by the classical equations of motion, the
Feynman path integrals still have to be solved. An exact solution
to those integrals is only available under certain approximations
such as, for example, the Gaussian approximation. That means that
only terms of order $0,1$ and $2$ in $\zeta, \xi, \dot{\zeta}$ and
$\dot{\xi}$ in the exponential can be  analytically computed.

Recognizing the necessity of making approximations in (\ref{ros}),
we will base ourselves upon physical reasoning to proceed further
with our calculations.

Firstly we can suppose that the interaction potential will be able
to attract the particles. While being attracted, they experience
the effect of friction until they end up very close to each other.
So, in thermal equilibrium we can assume that the reasonable
regime is that of short relative distances and the effective
potential can be considered harmonic with frequency $\omega_0$,
which has already been defined as $\omega_0^2 \equiv
4\eta\Omega/M\pi$.

Besides the potential $V_{eff}[\theta(\tau,\tau')]$,
the ``anomalous'' dissipation $\tilde{\eta}[\theta(\tau,\tau')]$ also has a functional
form that includes terms of order higher than $2$. We can again
assume the regime $|k_0\xi(\tau)| << 1$ and obtain an ohmic dissipation,
$\tilde{\eta}[\theta(\tau,\tau')]\approx \eta$.

These arguments transforms the formal solution in Eq.(\ref{ros}) into
an approximate solution that carries all the fundamental characteristics
and effects of the first one. Now the variables are decoupled and we can
write

\begin{equation}
\rho_S(\zeta_f,\zeta_i;\xi_f,\xi_i,\beta) = \rho_{\zeta}(\zeta_f,\zeta_i,\beta) \ \rho_\xi(\xi_f,\xi_i,\beta),
\end{equation}
where
\begin{equation}
\rho_{\zeta}(\zeta_f,\zeta_i,\beta) \equiv \int_{\zeta_i}^{\zeta_f}\mathcal{D}[\zeta(\tau)]\exp\left[ -\frac{1}{\hbar} S^{[1]}_{eff} \right]
\end{equation}
and
\begin{equation}
\rho_{\xi}(\xi_f,\xi_i,\beta) \equiv \int_{\xi_i}^{\xi_f}\mathcal{D}[\xi(\tau)]\exp\left[ -\frac{1}{\hbar} S^{[2]}_{eff} \right].
\end{equation}
In the equations above, we have
{\small\begin{equation}
S^{[1]}_{eff} \equiv \int_0^{\hbar\beta}\left[  M\dot{\zeta}^2 + 2\int_0^\tau d\tau'  K(\tau-\tau')(\zeta(\tau)-\zeta(\tau'))^2   \right]d\tau
\end{equation}}
and
{\small\begin{flushleft}
$S^{[2]}_{eff} \equiv$
\end{flushleft}
\begin{equation}
 \int_0^{\hbar\beta}\left[  \frac{M}{4}\dot{\xi}^2 + \frac{M}{4}\omega_0^2\xi^2 + \frac{1}{4}\int_0^\tau d\tau'  K(\tau-\tau')(\xi(\tau)-\xi(\tau'))^2   \right]d\tau.
\end{equation}}

The solution of the thermal equilibrium density matrix for a
Brownian particle in a harmonic potential can be found in
\cite{annals,Weiss}. This allows  us to write

\begin{equation}
\rho_{\zeta}(\zeta_f,\zeta_i,\beta) = C_\zeta \ \exp\left[   -\frac{1}{8\langle q_\zeta^2\rangle}(\zeta_f+\zeta_i)^2 -
\frac{\langle p_\zeta^2 \rangle}{2\hbar^2}(\zeta_f-\zeta_i)^2   \right]
\end{equation}
and
\begin{equation}
\rho_{\xi}(\xi_f,\xi_i,\beta) = C_\xi \ \exp\left[   -\frac{1}{8\langle q_\xi^2\rangle}(\xi_f+\xi_i)^2 -
\frac{\langle p_\xi^2 \rangle}{2\hbar^2}(\xi_f-\xi_i)^2   \right].
\end{equation}
The parameters of the equations are
\begin{equation}
\langle q_j^2 \rangle = \frac{2\hbar \gamma_j}{m_j\pi }\int_0^\Omega\left( \frac{\omega}{(\omega_j^2-\omega^2)^2+4\gamma_j^2\omega^2}  \right)
\coth\left( \frac{\hbar\omega\beta}{2} \right) d\omega
\label{qdois}\end{equation}
and
\begin{equation}
\langle p_j^2 \rangle = \frac{2m_j\hbar \gamma_j}{\pi }\int_0^\Omega\left( \frac{\omega^3}{(\omega_j^2-\omega^2)^2+4\gamma_j^2\omega^2}  \right)
\coth\left( \frac{\hbar\omega\beta}{2} \right) d\omega,
\label{pdois}\end{equation}
such that $\gamma_j \equiv \eta_j/(2m_j)$, $C_j$ is a normalization constant and $j = \zeta, \xi$.
The constants are given by

\begin{equation}
\omega_\zeta^2 = 0, \  m_\zeta = 2M  \ \mbox{and} \  \eta_\zeta = 4\eta
\label{const}\end{equation}
and
\begin{equation}
\omega_\xi^2 = \frac{4\eta\Omega}{M\pi},  \ m_\xi = \frac{M}{2} \ \mbox{and}  \  \eta_\xi = \frac{\eta}{2}.
\label{constdois}\end{equation}

Finally, we can return to the original variables of the problem, that is,
the coordinates of the individual particles. Then,

\begin{equation}
\rho_S(x_1,x_2;y_1,y_2,\beta) \equiv \langle x_1 \ x_2 |\rho_S|y_1 \ y_2\rangle =
\nonumber\eeq
\beq
 = C_\zeta C_\xi \exp[-f(x_1,x_2,y_1,y_2)],
\label{rororo}\end{equation}
where
\begin{equation}
f(x_1,x_2,y_1,y_2)   \equiv   \frac{(y_1-y_2+x_1-x_2)^2}{8\langle q_\xi^2\rangle} +
\nonumber\eeq
\beq
+ \frac{\langle p_\xi^2\rangle}{2\hbar^2}
(y_1-y_2-x_1+x_2)^2
 +\frac{(y_1+y_2+x_1+x_2)^2}{32\langle q_\zeta^2\rangle} +
\nonumber\eeq
\beq
+ \frac{\langle p_\zeta^2\rangle}{8\hbar^2}
(y_1+y_2-x_1-x_2)^2.
\end{equation}
So, we have (\ref{rororo}) as the thermal equilibrium density
matrix for two Brownian particles in a common heat bath. In the
next section we evaluate a measure of entanglement in such state
as a function of the temperature of the system and of the coupling
between the particles and the reservoir.

\section{III. Thermal entanglement between two quantum brownian particles}

At this stage we are able to answer the question whether it is
possible for the bath to create or  keep the entanglement between
two open quantum systems in thermal equilibrium with it. We
describe the entanglement as a function of the temperature of the
system and the coupling between the particles and the reservoir.
The latter manifests itself through the dependence of the
entanglement on the damping constant.

We then conclude that in fact the effective potential mediated by
the bath is responsible  for the state non-separability.
Furthermore, we see that there is a range of finite temperatures
for which this property persists, no matter what the initial
states of the two particles is.

A general Gaussian state $\rho$ can have its entanglement suitably
quantified by the Logarithmic Negativity, $E_{\mathcal{N}}[\rho]$
\cite{Adesso}. We see by (\ref{rororo}) that indeed it is the case
of the state $\rho_S$.

We now make a brief review of the main aspects of this measure.
First of all, it is worth to say that it is based on a criterion of separability
proposed by R. Simon \cite{Simon}, who generalized the ideas of A. Peres \cite{Peres}
about {\it positive partial transposition} violation.

A two mode Gaussian state can be written in terms of a characteristic function, defined by

\begin{equation}
\tilde{W}(X) = \exp\left[ -\frac{1}{2}X \ \sigma \ X^T \right],
\label{funccarac}\end{equation}
where $X \equiv (q_1, p_1, q_2, p_2)$ and $\sigma$ is the covariance matrix \cite{Adesso}.
The characteristic function relates to the density matrix through the relation

\begin{equation}
\tilde{W}(X) = \int_{-\infty}^{\infty} e^{-i \vec{p}.\vec{r}} \ \left\langle  \vec{r}-\frac{\vec{q}}{2} \right| \rho_S \left|
\vec{r}+\frac{\vec{q}}{2} \right\rangle d^2r.
\label{wdero}\end{equation}

Eqs.(\ref{funccarac}) and (\ref{wdero}) connect the state of the
system with its covariance matrix and that is basically the reason
for their introduction. The interest in the covariance matrix is a
consequence of the fact that it caries all the information about
the separability of the state. Its finite dimension  is
fundamental for the applicability of the referred criteria
\cite{Peres, Simon}.

The general form of a two mode covariance matrix is

\begin{equation}
\sigma = \left(
\begin{array}{cc}
\alpha & \gamma \\
\gamma^T & \beta
\end{array}
\right),
\end{equation}
where $\alpha, \beta$ and $\gamma$ are $2\times 2$ matrices. With them, we define the local sympletic invariants

\begin{equation}
\tilde{\Delta} \equiv \det \alpha + \det \beta - 2\det \gamma
\end{equation}
and
\begin{equation}
2 \tilde{\nu}_{-}^2 \equiv \tilde{\Delta} - \sqrt{\tilde{\Delta}^2 - 4 \det\sigma}.
\end{equation}
The new parameter $\tilde{\nu}_{-}$ is the so called sympletic
eigenvalue of the partial transposed density matrix and is related
to the logarithmic negativity \cite{Adesso} through
\begin{equation}
E_{\mathcal{N}}[\rho_S] = \mbox{max}[0,\ -\ln\tilde{\nu}_{-}].
\end{equation}

After some algebra, we show that for the density matrix in Eq.(\ref{rororo})
we have

\begin{equation}
\tilde{\nu}_{-} = \frac{\sqrt{\langle p_\zeta^2\rangle\langle q_\xi^2\rangle}}{2\hbar},
\end{equation}
which implies that the entanglement is given by

\begin{equation}
E_{\mathcal{N}}[\rho_S] = \mbox{max}\left[0, \ -\frac{1}{2}\ln\left( \frac{\langle p_\zeta^2\rangle\langle q_\xi^2\rangle}{4\hbar^2} \right)\right].
\label{emaranhamento}\end{equation}

To plot this function with respect to the temperature of the
system we define, with the help of (\ref{constdois}), the
dimensionless variables

\begin{equation}
r \equiv \frac{\gamma_\zeta}{\omega_0}  \ \ \mbox{and} \  \  A \equiv \left( \frac{k_B T}{\hbar\gamma_\zeta} \right)^2.
\end{equation}
The parameter $r$ measures the coupling between the particle and
the reservoir. The parameter $A$ measures the temperature. In the
low temperature ($0 < A < 1$) and underdamped ($0 < r < 1$)
limits, the entanglement behaves as in Fig.(1).

\begin{figure}
\begin{centering}
\includegraphics{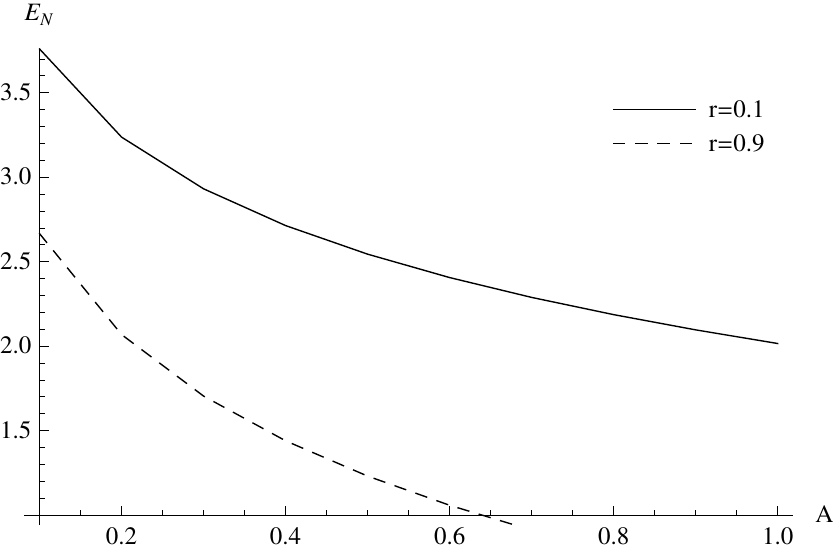}
\par\end{centering}

\caption{\label{fig: Enmed1}Logarithm Negativity as a function of the
temperature $A$. The coupling between the system and the heat bath is
parametrized by $r$. The graphic is valid for the low temperature and
underdamping limit. Observe that the origin of the graphic is
($0.1, 1.0$).}

\end{figure}

As expected, the entanglement between the particles  monotonically
decreases with the increase of temperature. The other significant
point is that the higher the coupling between the system and the
reservoir, the faster the entanglement is lost with the increase
of temperature.

\section{IV. Conclusions}

The focus of this work was the study of the thermal equilibrium
quantum properties of two Brownian particles in a common
reservoir. More specifically, we evaluated the equilibrium density
matrix for the particles and the entanglement between them.

The presented model was based on the system-plus-reservoir
approach. The excitations of the bath were described by harmonic
oscillators and the coupling was nonlinear in the system
coordinates. The separation of the length scale from the time
scale in the response function allowed us to analytically study an
effective interaction between the particles, which is responsible
for the entanglement in their state. The chosen response function
reproduces the Brownian motion for each particle when they are far apart.

We obtained a density matrix that, in terms of Feynman path
integrals, presents, besides the effective potential,  the
``anomalous dissipation'' explicitly in its functional form. The
center of mass behaved like a free quantum Brownian particle in
thermal equilibrium, with a friction dependent on the distance
between the particles. The relative coordinate behaved like a
quantum Brownian particle in a binding potential, in thermal
equilibrium.

Finally, we concluded that the environment was not only
responsible for the loss of coherence, as intuitively expected,
but also for the induction of entanglement. Even for finite
temperatures this phenomenon still persists, independent of the
initial conditions, since only thermal equilibrium was considered.
The entanglement decreased monotonically with the increase of
temperature. Such a decrease was enforced by the increase of the
coupling between the system and the environment.

In our view the establishment of entanglement between those two
particles mediated by the bath reflects the existence of a complex
ground state of the composite system which in general is far from
being separable.

Our results may be relevant for a better understanding of quantum
information in non-isolated systems. They can also be applied to
other configurations of the system of interest, for example, the
one in which each particle is in an independent potential, again
in the same bath. This model may be also adaptable to other
realistic systems, opening the possibility to study quantum
properties of general continuous variable open systems.

\section{Acknowledgments}

We kindly acknowledge Funda\c{c}{\~a}o de Amparo {\`a} Pesquisa do
Estado de S{\~a}o Paulo (FAPESP) and the Conselho Nacional de
Desenvolvimento Cient{\'\i}fico e Tecnol{\'o}gico (CNPq) for
financial support. AOC also thanks the support from the Instituto
Nacional de Ci{\^e}ncia e Tecnologia em Informa\c{c}{\~a}o
Qu{\^a}ntica (INCT-IQ).

\end{document}